\documentclass[twocolumn,showpacs,floatfix,prl]{revtex4}

\usepackage[dvips]{epsfig}

\usepackage{float}

\begin{document}

\title{$Z_2$ Topological Order and the Quantum Spin Hall Effect}

\author{C.L. Kane and E.J. Mele}
\affiliation{Dept. of Physics and Astronomy, University of Pennsylvania,
Philadelphia, PA 19104}

\begin{abstract}
The quantum spin Hall (QSH) phase is a time reversal invariant
electronic state with a bulk electronic band gap that supports the
transport of charge and spin in gapless edge states.
We show that this phase is associated with a
novel $Z_2$ topological invariant, which distinguishes
it from an ordinary insulator.  The $Z_2$ classification, which is defined
 for time reversal invariant Hamiltonians, is analogous to the Chern number
classification of the quantum Hall effect.  We establish the $Z_2$
order of the QSH phase in the two band model of graphene and propose a
generalization of the formalism applicable to multi band and interacting systems.

\end{abstract}

\maketitle

The classification of electronic states according to
topological invariants is a powerful tool for understanding many
body phases which have bulk energy gaps.  This approach was pioneered by Thouless et
al.\cite{tknn} (TKNN), who identified the topological invariant for
the non interacting integer quantum Hall effect.  The TKNN integer,
$n$, which gives the quantized Hall conductivity for each band $\sigma_{xy}=n e^2/h$,
is given by an integral of the Bloch wavefunctions
over the magnetic Brillouin zone, and corresponds to the
first Chern class of a $U(1)$ principal fiber bundle on a torus\cite{avron,kohmoto}.
An equivalent formulation, generalizable to interacting
systems, is to consider the sensitivity of the many body ground state
to phase twisted periodic boundary conditions\cite{niu1,arovas}.
%In this case the Brillouin zone
%integral is replaced by an integral of the many body wavefunction
%over the two dimensional torus of boundary condition phases.
This topological classification distinguishes a simple insulator
%(with TKNN number zero)
from a quantum Hall state, and
explains the insensitivity of the Hall conductivity to weak disorder
and interactions.  Nonzero TKNN integers are also intimately
related to the presence of gapless edge states on the sample
boundaries\cite{hatsugai}.

Since the Hall conductivity violates time reversal (TR) symmetry, the TKNN
integer must vanish in a TR invariant system.  Nonetheless,
we have recently shown that the spin orbit interaction in a single plane of
graphene leads to a TR invariant quantum spin Hall
(QSH) state which has a bulk energy gap, and a pair of
gapless spin filtered edge states on the boundary\cite{km}.
In the simplest version of our model (a $\pi$ electron tight binding model
with mirror symmetry about the plane)
the perpendicular component of the spin, $S_z$,
is conserved.  Our model then reduces to independent copies for each
spin of a model introduced by Haldane\cite{haldane}, which exhibits an
integer quantum Hall effect even though the average magnetic field is
zero.  When $S_z$ is conserved the
distinction between graphene and a simple insulator
is thus easily understood.   Each spin
has an independent TKNN integer $n_\uparrow$, $n_\downarrow$.
TR symmetry requires $n_\uparrow+n_\downarrow=0$, but the
difference $n_\uparrow-n_\downarrow$ is nonzero and defines a
quantized spin Hall conductivity.

This characterization breaks down when $S_z$ non conserving terms are
present.  Such terms will inevitably arise due to coupling to other
bands, mirror symmetry breaking terms, interactions or disorder.
Though these perturbations destroy the quantization of the spin Hall
conductance, we argued that they do not destroy the topological order
of the QSH state because Kramers' theorem prevents TR
invariant perturbations from opening a gap at the edge\cite{km}.  Thus, even
though the single defined TKNN number (the total Hall conductance) is
zero, the QSH groundstate is distinguishable from that of a
simple insulator.  This suggests that there must be an additional
topological classification for TR invariant systems.

In this paper we clarify the topological order of the QSH phase and
introduce a $Z_2$ topological index that characterizes
TR invariant systems.  This classification is similar
to the TKNN classification, and gives a simple
test which can be applied to Bloch energy bands to distinguish the
insulator from the QSH phase.  It
may also be formulated as a
sensitivity to phase twisted boundary conditions.
We will begin by describing our model of
graphene and demonstrate that the QSH phase is robust even when
$S_z$ is not conserved.  We will then analyze the
constraints of TR invariance and derive the $Z_2$ index.

Consider the tight binding Hamiltonian of graphene introduced in Ref. \onlinecite{km},
which generalizes Haldane's model\cite{haldane}
 to include spin with TR invariant spin
orbit interactions.
\begin{eqnarray}
H=t\sum_{\langle ij\rangle} c_i^\dagger c_j
+ i\lambda_{SO} \sum_{\langle\langle ij \rangle\rangle}\nu_{ij} c_i^\dagger s^z c_j \\
+ i\lambda_R\sum_{\langle ij\rangle} c_i^\dagger
({\bf s}\times\hat{\bf d}_{ij})_z c_j
 + \lambda_v\sum_i  \xi_i c_i^\dagger c_i.\nonumber
\end{eqnarray}
The first term is a nearest neighbor hopping term
on the honeycomb lattice, where we have suppressed the spin index on the
electron operators.  The second term is the mirror symmetric spin
orbit interaction which involves spin dependent second neighbor hopping.
Here $\nu_{ij} = (2/\sqrt{3})(\hat {\bf
d}_1\times \hat{\bf d}_2)_z = \pm 1$, where $\hat{\bf d}_1$ and $\hat{\bf d}_2$
are unit vectors along the two bonds the electron traverses going
from site $j$ to $i$.  $s^z$ is a Pauli matrix
describing the electron's spin.  The third term is a nearest
neighbor Rashba term, which explicitly violates the $z \rightarrow -z$
mirror symmetry, and will arise due to a perpendicular electric field
or interaction with a substrate.  The fourth term is a staggered
sublattice potential ($\xi_i = \pm 1$), which we include to describe the
transition between the QSH phase and the simple insulator.
This term violates the symmetry under twofold rotations in the plane.

$H$ is diagonalized by writing
$\phi_s({\bf R}+\alpha{\bf d}) = u_{\alpha s}({\bf k}) e^{i{\bf k}\cdot{\bf R}}$.
Here $s$ is spin and ${\bf R}$ is a bravais lattice vector built from primitive
vectors ${\bf a}_{1,2}=(a/2)(\sqrt{3}\hat y \pm \hat x)$.  $\alpha=0,1$ is
%=n_1 {\bf a}_1+n_2 {\bf a}_2
the sublattice index with ${\bf d}=a \hat y/\sqrt{3}$.
%Note that with this definition
%$u_{as}({\bf k}+{\bf G}) = u_{as}({\bf k})$.
For each ${\bf k}$ the Bloch
wavefunction is a four component eigenvector $|u({\bf k})\rangle$ of
the Bloch Hamiltonian matrix ${\cal H}({\bf k})$.
The 16 components of ${\cal H}({\bf k})$
may be written in terms of the identity matrix, 5 Dirac matrices $\Gamma^a$
and their 10 commutators $\Gamma^{ab}=[\Gamma^a,\Gamma^b]/(2i)$\cite{murakami1}.
We choose the following representation of the Dirac matrices:
$\Gamma^{(1,2,3,4,5)} = (\sigma^x\otimes I,\sigma^z\otimes I,\sigma^y\otimes s^x,
\sigma^y\otimes s^y, \sigma^y \otimes s^z$), where the Pauli matrices
$\sigma^k$ and $s^k$ represent the sublattice  and spin indices.
%representations are also possible,
This choice organizes the matrices according to TR.  The TR operator is given by
$\Theta |u\rangle \equiv  i (I\otimes s^y)|u\rangle^*$.
The five Dirac matrices are even under TR,
$\Theta\Gamma^a\Theta^{-1} = \Gamma^a$
%$\bar \Gamma^a \equiv (I \otimes s^y) (\Gamma^a)^* (I\otimes s_y) =
%\Gamma^a$,
while the 10 commutators are odd,
%$\bar\Gamma^{ab} = -\Gamma^{ab}$.
$\Theta\Gamma^{ab}\Theta^{-1}=-\Gamma^{ab}$.
The Hamiltonian is thus
\begin{equation}
{\cal H}({\bf k}) = \sum_{a=1}^5 d_a({\bf k}) \Gamma^a +
\sum_{a<b=1}^5
d_{ab}({\bf k})
\Gamma^{ab}
\end{equation}
where the $d({\bf k})$'s are given in Table I.
%, with
%$x = k_x a/2$ and $y = \sqrt{3} k_y a/2$.
%$d_1= t(1+2 \cos x\cos y)$,
%$d_2=\lambda_v$,
%$d_3=\lambda_R(1-\cos x\cos y)$,
%$d_4=-\sqrt{3}\lambda_R \sin x\sin y$,
%$d_{12}= -2 t \cos x\sin y$,
%$d_{15}= \lambda_{SO}(2\sin 2x - 4\sin x \cos y)$
%$d_{23}=-\lambda_R\cos x\sin y$,
%$d_{24}=\sqrt{3}\lambda_R\sin x\cos y$ (here
% The remaining 7 d's are zero.
Note that ${\cal H}({\bf k}+{\bf G})={\cal H}({\bf k})$
for reciprocal lattice vectors ${\bf G}$,
so ${\cal H}({\bf k})$ is defined on a torus.
The TR invariance of
${\cal H}$ is reflected in the symmetry (antisymmetry) of $d_a$
$(d_{ab})$ under ${\bf k}\rightarrow -{\bf k}$.
\begin{table}
  \centering
  \begin{tabular}{|cc|cc|} \hline
  % after \\ : \hline or \cline{col1-col2} \cline{col3-col4} ...
$d_1$ & $t(1+2 \cos x \cos y)$ & $ d_{12}$ & $-2t \cos x \sin y$ \\
$d_2$ & $\lambda_v$ & $d_{15}$ & $\lambda_{SO} ( 2 \sin 2x - 4 \sin x \cos
y)$  \\
$d_3$ & $\lambda_R (1 - \cos x \cos y)$ & $d_{23}$ & $-\lambda_R \cos x
\sin y$ \\
$d_4$ & $-\sqrt{3} \lambda_R \sin x \sin y $ & $d_{24}$ & $\sqrt{3}
\lambda_R \sin x \cos y $\\
  \hline
\end{tabular}
  \caption{The nonzero coefficents in Eq. 2 with $x=k_x a/2$ and $y=\sqrt{3}k_y a/2$.}
\end{table}

For $\lambda_R=0$ the there is an energy gap with magnitude
$|6\sqrt{3}\lambda_{SO}-2\lambda_v|$.
For $\lambda_v> 3\sqrt{3}\lambda_{SO}$ the  gap is dominated by
$\lambda_v$, and the system is an insulator.
$3\sqrt{3}\lambda_{SO} >\lambda_v$ describes the QSH phase.
Though the  Rashba term violates $S_z$ conservation,
for $\lambda_R< 2\sqrt{3}\lambda_{SO}$ there is a
finite region of the phase diagram in Fig. 1 that is adiabatically
connected to the QSH phase at $\lambda_R=0$.  Fig. 1 shows
the energy bands obtained by solving the lattice model in a zigzag strip
geometry\cite{km} for representative points in the
insulating and QSH phases.  Both phases have a bulk energy
gap and edge states, but in the QSH phase the edge states traverse
the energy gap in pairs.   At the transition between the two phases, the energy
gap closes, allowing the edge states to ``switch partners".

\begin{figure}
 \centerline{ \epsfig{figure=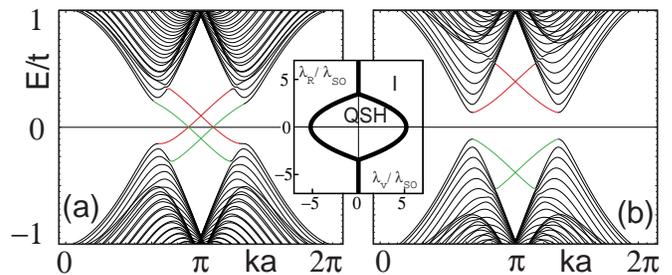,width=3.4in} }
 \caption{Energy bands for a one dimensional ``zigzag" strip in the (a) QSH phase
$\lambda_v=.1 t$ and (b) the insulating phase $\lambda_v=.4 t$.  In both cases
 $\lambda_{SO}=.06 t$ and $\lambda_R=.05 t$.  The edge states on a given edge
 cross at $ka = \pi$.  The inset shows the phase diagram as a function of
 $\lambda_v$ and $\lambda_R$ for $0<\lambda_{SO}\ll t$. }
 \end{figure}

The behavior of the edge states signals a clear difference between
the two phases.  In the QSH phase for each energy in the bulk gap
there is a single time reversed pair of eigenstates on each edge.
Since TR symmetry prevents the mixing of Kramers' doublets
these edge states are robust against small perturbations.  The
gapless states thus persist even if the spatial symmetry is further
reduced (for instance by removing the $C_3$ rotational symmetry in
(1)).  Moreover, weak disorder will not lead to localization of the
edge states because single particle elastic backscattering is
forbidden\cite{km}.

In the insulating state the edge states do not traverse
the gap.  It is possible that for certain edge potentials the edge
states in Fig. 1b could dip below the band edge, reducing - or even
eliminating - the edge gap.  However, this is still distinct from the
QSH phase because there will necessarily be an {\it even} number of
Kramers pairs at each energy.  This allows elastic backscattering, so
that these edge states will in general be localized by weak disorder.
The QSH phase is thus distinguished from the simple insulator by the
number of edge state pairs modulo 2.  Recently two dimensional
versions\cite{zhang} of the spin Hall insulator models\cite{murakami2}
have been introduced,
which under conditions of high spatial symmetry exhibit gapless edge
states.  These models, however, have an even number of edge
state pairs.  We shall see below that they are topologically
equivalent to simple insulators.

The QSH phase is {\it not} generally characterized by a quantized spin Hall
conductivity.
%A number of authors have focused on the spin Hall
%conductivity computed via the Kubo formula\cite{murakami1,murakami2}.
%When spin is not conserved the Kubo spin Hall
%conductivity need not be quantized, and will be non zero even in a simple
%insulator.  However the meaning of this is not clear because  spin currents can not be
%directly measured.  We will instead focus on
Consider the rate of spin accumulation
at the opposite edges of a cylinder of circumference $L$,
which can be computed using Laughlin's
argument\cite{laughlin}.
% Consider a cylinder with circumference $L$ and two
%edges at opposite ends.
A weak circumferential electric field $E$ can be induced by adiabatically
threading
magnetic flux through the cylinder.  When the flux increases
by $h/e$ each momentum eigenstate shifts by one unit: $k
\rightarrow k+2\pi/L$.  In the insulating state (Fig. 1b)
this has no effect, since the valence band is
completely full.  However, in the QSH state a particle-hole excitation
is produced at the Fermi energy $E_F$.  Since
the particle and hole states do not have the
same spin, spin accumulates at the edge.  The rate of spin
accumulation defines a spin Hall conductance $d\langle S_z \rangle/dt
= G^s_{xy}E$, where
\begin{equation}
G^s_{xy}={e\over h}\left(
\langle S_z \rangle_L -\langle S_z
\rangle_R\right)|_{E_F}.
\end{equation}
Here the expectation value of $S_z$ is evaluated for the left and
right moving states at $E_F$.
Since the edge states
are not necessarily $S_z$ eigenstates this spin Hall conductance is
not quantized.
$G^s_{xy}$ is zero in the insulating phase,
though,  provided $E_F$ is in the gap at the edge.
If in the insulator the edge states cross $E_F$, then in a
clean system there could be spin accumulation at the edge (resulting from the
acceleration of the edge electrons in response to $E$).
However, if the edge states are localized then there will be
no spin accumulation.  Thus the nonzero spin accumulation persists
only for the QSH phase, justifying the term quantum
(but not quantized) spin Hall effect.

In the quantum Hall effect, the states with zero and one flux quantum
threading the cylinder are distinguished by the charge
polarization.  The two states can not be connected by any
operator that locally conserves charge.
In the QSH effect there is no simple conserved quantity
 distinguishing the two states.  However, the states {\it are}
distinguishable, because the state with an edge particle-hole excitation at
$E_F$ can not be connected to the ground state with a
local TR symmetric operator.  Note, however, that if a
second flux is added, then there will be TR invariant
interactions which do connect the state with the zero flux state.  This
suggests that the state with one flux added is distinguished by a
$Z_2$ ``TR polarization".

The classification of quantum Hall states on the
cylinder according to Laughlin's argument is intimately related to
the TKNN classification of the Bloch wavefunctions\cite{niu1}.  To
establish the corresponding topological classification for TR
invariant systems we consider TR constraints
 on the set of occupied Bloch wavefunctions
$|u_{i=1,2}({\bf k})\rangle$.  $|u_i({\bf k})\rangle$ form a rank 2 vector bundle over
%the
%
%on the mapping from the Brillouin zone to the occupied Bloch
%wavefunctions.  The set of occupied Bloch wavefunctions $|u_i\rangle$
%form a rank 2 vector bundle over the
Brillouin zone torus.  TR introduces an involution on the
torus which identifies pairs of points ${\bf k}$ and ${\bf -k}$.
Wavefunctions at the identified points are related by $|u_i(-{\bf
k})\rangle=\Theta |u_i({\bf k})\rangle$, implying
that the bundle is ``real".  Since $\Theta^2=-1$, $\Theta$ has period
4, so that the real bundle is ``twisted".  These bundles are classified within the
mathematical framework of twisted Real K theory\cite{atiyah1}.  It is found that
such bundles have a $Z \times Z_2$ classification on a torus\cite{thanks}.
The first integer
gives the rank of the bundle (i.e. the number of occupied bands).
The $Z_2$ index is related to the mod 2 index of the real Dirac
operator\cite{atiyah2}.   In the following we will explicitly construct this $Z_2$
index from the Bloch wavefunctions and show that it distinguishes the
QSH phase from the simple insulator.

%For each value of ${\bf k}$ the ground state wavefunction consists of a
%Slater determinant of the four component
%wavefunctions $|u_1\rangle$ and $|u_2\rangle$ of the two occupied bands.
%The manifold of possible ground states may be identified by noting that a set of four
%orthonormal complex eigenvectors identify an element of $U(4)$.  Since
%the ground state is unchanged by arbitrary $U(2)$ rotations among
%either the occupied or unoccupied states the distinct states are
%elements of $U(4)/(U(2)\times U(2))$, also known as the complex
%Grassmanian manifold $G_{4,2}({\cal C})$.  The dimension of this manifold is
%8. This can be seen by noting that the complex vectors $u_1\rangle$ and
%$u_2\rangle$ are specified by 16 real parameters.  Orthonormality
%gives 4 constraints, and there is a 4 parameter $U(2)$ redundancy.

TR symmetry identifies two important subspaces of the
space of Bloch Hamiltonians ${\cal H}({\bf k})$ and the corresponding
occupied band wavefunctions $|u_i({\bf k})\rangle$.  The ``even" subspace,
for which $\Theta H \Theta^{-1}=H$ have wavefunctions with the property that
$\Theta|u_i\rangle$ is equivalent to $|u_i\rangle$ up to a
$U(2)$ rotation.  From Eq. 2 it
is clear that in this subspace $d_{ab}({\bf k})=0$.
% Thus, the groundstates
%may be identified with the surface $S^4$ of a 5 dimensional sphere.
TR symmetry requires that $H({\bf k})$
 belong to the even subspace at the $\Gamma$ point ${\bf
k}=0$ as well as the three $M$ points shown in Fig. 2a,b.
The odd subspace has
wavefunctions with the property that the space spanned by $\Theta|u_i({\bf k})\rangle$
is {\it orthogonal} to the space spanned by $|u_i({\bf k})\rangle$.
We shall see that the $Z_2$ classification may be found by studying the
set of ${\bf k}$ which belong to the odd subspace.

The special subspaces can be identified by considering the matrix
of overlaps,
$\langle u_i({\bf k})|\Theta|u_j({\bf k})\rangle$.
From the properties of $\Theta$ it is clear that this matrix is
antisymmetric, and may be expressed in terms of a single complex
number as  $\epsilon_{ij} P({\bf k})$.  $P({\bf k})$ is in fact equal
to the {\it Pfaffian}
\begin{equation}
P({\bf k}) = {\rm Pf} \left[\langle u_i({\bf k})|\Theta|u_j({\bf
k})\rangle\right],
\end{equation}
which for a 2 by 2 antisymmetric matrix $A_{ij}$ simply picks out $A_{12}$.
We shall see below that the Pfaffian is the natural
generalization when there are more than two occupied bands.
$P({\bf k})$ is not gauge invariant.  Under a $U(2)$
transformation $|u'_i\rangle = U_{ij}|u_j \rangle$,
$P' = P\det U $.  Thus $P$ is unchanged by a $SU(2)$ rotation, but under a
$U(1)$ transformation $U = e^{i\theta}$, $P' = P e^{2 i\theta}$.
In the even subspace $\Theta|u_i\rangle$ is equivalent to $|u_i\rangle$ up to
a $U(2)$ rotation, and we have $|P({\bf k})|=1$.  In
the odd subspace $P({\bf k})=0$.

If no spatial symmetries constrain its form, the
zeros of $P({\bf k})$ are found by tuning two parameters,
and generically occur at points in the Brillouin zone.
First order zeros occur at time reversed pairs of points
$\pm {\bf k}^*$ with opposite ``vorticity", where the
phase of $P({\bf k})$ advances in opposite directions around $\pm{\bf k}^*$.
For $\lambda_v\ne 0$ the QSH phase is distinguished from the
simple insulator by the presence of a single pair of first order zeros of $P({\bf k})$.
The $C_3$ rotational symmetry of our model constrains ${\bf k}^*$ to be at
the corner of the Brillouin zone as shown in Fig 2a.
If the $C_3$ symmetry is relaxed, ${\bf k}^*$ can occur anywhere
{\it except} the four symmetric points where $|P({\bf k})|=1$.  The number of
pairs of zeros is a $Z_2$ topological invariant.  This can be seen
by noting that two pairs $\pm {\bf k}^*_{1,2}$
can come together to annihilate
each other when ${\bf k}_1^* = - {\bf k}_2^*$.
However a single pair of zeros at $\pm {\bf k}^*$ can not annihilate
because they would have to meet at either $\Gamma$ or $M$,
where $|P({\bf k})|=1$.
If TR symmetry is broken then the zeros
are no longer prevented from annihilating, and the topological distinction of
the QSH phase is lost.

The $Z_2$ index can thus be determined by counting the number of
pairs of complex zeros of $P$.  This can be accomplished by
evaluating the winding of the phase of $P({\bf k})$ around a loop
enclosing {\it half} the Brillouin zone (defined so that ${\bf k}$ and $-{\bf k}$
are never both included).
\begin{equation}
I ={1\over{ 2\pi i}}\oint_C d{\bf k}\cdot\nabla_{\bf k}\log (P({\bf
k})+ i\delta),
\end{equation}
where $C$ is the path shown in Fig. 2a,b.

\begin{figure}
 \centerline{ \epsfig{figure=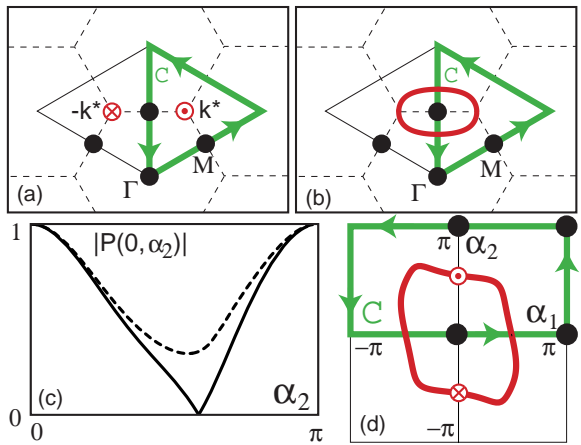,width=3in} }
 \caption{The zeros of $P({\bf k})$ in the QSH phase occur at points
 $\pm{\bf k}^*$ for (a) $\lambda_v\ne 0$ and on the oval for
 (b)  $\lambda_v=0$.  (c) $|P(0,\alpha_2)|$ in the QSH (solid)
and insulating (dashed) phases for a $2\times 2$ supercell
using parameters in Fig. 1.
 (d) Point ($\lambda_v\ne 0$) and line($\lambda_v=0$) zeros of
 $P({\vec\alpha})$ for the $2\times 2$ supercell.
In (a,b,d) the solid dots are
 TR symmetric points, which can't be zeros of $P$, and
$C$ is the contour of integration for Eq. (5).
 }
 \end{figure}

When $\lambda_v=0$ (as it is in graphene)
$H$ has a $C_2$ rotational symmetry, which when combined with TR
constrains the form of ${\cal H}({\bf k})$, and allows
$P({\bf k})$ to be chosen to be real.
The zeros of $P({\bf k})$ then occur along lines,
rather than at points.  We find that the zeros are absent in the
insulating phase, but enclose the $M$ point in the
QSH phase as shown in Fig. 2b.  In this case we find that
Eq. 4 also determines the $Z_2$ index (given by 1/2 the number of sign changes
along the path $C$), provided we include the
convergence factor $\delta$.  Note that though the sign of $I$ depends on the
sign of $\delta$, $I$ mod 2 does not.  We thus conclude that the QSH
phase and the insulator are distinguished by the $Z_2$ index $I$.

The spin Hall insulator models studied in Refs. \onlinecite{zhang,murakami2} are
simple insulators with $I=0$.  Their Hamiltonian,
when expressed in the form of Eq. 2, has $d_{ab}({\bf k})=0$, so that
$|u_i({\bf k})\rangle$ is in the even subspace and $|P({\bf k})|=1$ for all ${\bf k}$.
Ref. \onlinecite{zhang2} introduces a model which does appear to
exhibit a QSH effect.

Having established the topological classification of the Bloch
wavefunctions we now ask whether, in analogy with Ref. \onlinecite{niu1} the
classification can be formulated in terms of the sensitivity of the ground state
wavefunction to phase twisted periodic boundary conditions.
Such a formulation will address the topological
stability of the many body groundstate with respect to weak disorder and electron
interactions.  It also
provides the appropriate generalization of (4,5) for multi band
Hamiltonians.
Consider a ${\bf L}_1 \times {\bf L}_2$ sample with boundary condition
$\Psi(...,{\bf r}_i+{\bf L}_k,...) = e^{i\alpha_k}\Psi(...,{\bf r}_i,...)$.  For
concreteness we consider a rectangular geometry, with ${\bf L}_1 =
N_1 ({\bf a}_1+{\bf a}_2)$ and ${\bf L}_2 = N_2 ({\bf a}_1-{\bf
a}_2)$.  For non interacting electrons, we may view the entire sample
as a large unit cell with $N_a = 4 N_1 N_2$ atoms imbedded in an even
larger crystal.  Then $\vec\alpha$ plays the role of
${\bf k}$, and the occupied single particle eigenstates
$\phi_i(\vec\alpha)$ play the role of $u_i({\bf k})$.  $\phi_i(\vec\alpha)$
form a rank $N_a$ bundle on the torus defined by
$\alpha_{1,2}$.  The $Z_2$ classification can be obtained by
studying the zeros of
\begin{equation}
P(\vec \alpha) = {\rm Pf}\left[\langle
\phi_i(\vec\alpha)|\Theta|\phi_j(\vec\alpha)\rangle\right].
\end{equation}
Fig. 2c compares $|P(\vec\alpha)|$ in the QSH and insulating phases.
for a 16 site sample with $N_1=N_2=2$.
In the insulating phase there are no zeros.  In  the QSH phase
the structure of the zeros in Fig. 2d is similar to Fig. 2a,b.
For $\lambda_v\ne 0$ the first order zeros are at points, while
for $\lambda_v=0$ they are on a loop.  The zeros can not be at the
four TR symmetric points.  This
structure persists in the QSH phase for any cell size.
The $Z_2$ index $I$ can be computed by performing the integral
analogous to (5) along the contour $C$ in Fig. 2d.

A many body formulation requires the index to be expressed in terms
of the many particle groundstate $|\Phi(\vec\alpha)\rangle$.  It is
interesting to note that for non interacting electrons
 $\langle\Phi(\vec\alpha)|\Theta|\Phi(\vec\alpha)\rangle= {\rm
 det}[\langle\phi_i(\vec\alpha)|\Theta|\phi_j(\vec\alpha)\rangle] =
 P(\vec\alpha)^2$.  This suggests a many body generalization
 \begin{equation}
 P(\vec\alpha) =
 \sqrt{\langle\Phi(\vec\alpha)|\Theta|\Phi(\vec\alpha)\rangle}.
 \end{equation}
We suspect that with this definition
 the topological structure  $P(\vec\alpha)$ in Fig. 2c,d will remain in the
 presence of weak electron interactions.

 To conclude, we have introduced a $Z_2$ topological classifiation
 of TR invariant systems, analogous to the TKNN
 classification of quantum Hall states.  This shows that the QSH
 phase of graphene has a topological stability that is
 insensitive to weak disorder and interactions.

We thank Tony Pantev for many helpful discussions.
This
work was supported by the NSF under MRSEC grant DMR-00-79909 and the
DOE under grant DE-FG02-ER-0145118.


\begin{thebibliography}{10}
\bibitem{thanks}  We are indebted to Tony Pantev for explaining these
connections to us.
\end{thebibliography}
\end{document}